\documentclass[10pt]{article}

\begin{document}

\def\CA{{\cal A}}
\def\CB{{\cal B}}
\def\CC{{\cal C}}
\def\CD{{\cal D}}
\def\CE{{\cal E}}
\def\CF{{\cal F}}
\def\CG{{\cal G}}
\def\CH{{\cal H}}
\def\CI{{\cal I}}
\def\CJ{{\cal J}}
\def\CK{{\cal K}}
\def\CL{{\cal L}}
\def\CM{{\cal M}}
\def\CN{{\cal N}}
\def\CO{{\cal O}}
\def\CP{{\cal P}}
\def\CQ{{\cal Q}}
\def\CR{{\cal R}}
\def\CS{{\cal S}}
\def\CT{{\cal T}}
\def\CU{{\cal U}}
\def\CV{{\cal V}}
\def\CW{{\cal W}}
\def\CX{{\cal X}}
\def\CY{{\cal Y}}
\def\CZ{{\cal Z}}

\newcommand{\todo}[1]{{\em \small {#1}}\marginpar{$\Longleftarrow$}}
\newcommand{\labell}[1]{\label{#1}\qquad_{#1}} 
\newcommand{\bbibitem}[1]{\bibitem{#1}\marginpar{#1}}
\newcommand{\llabel}[1]{\label{#1}\marginpar{#1}}

\newcommand{\sphere}[0]{{\rm S}^3}
\newcommand{\su}[0]{{\rm SU(2)}}
\newcommand{\so}[0]{{\rm SO(4)}}
\newcommand{\bK}[0]{{\bf K}}
\newcommand{\bL}[0]{{\bf L}}
\newcommand{\bR}[0]{{\bf R}}
\newcommand{\tK}[0]{\tilde{K}}
\newcommand{\tL}[0]{\bar{L}}
\newcommand{\tR}[0]{\tilde{R}}

\newcommand{\ack}[1]{[{\bf Ack!: {#1}}]}

\newcommand{\btzm}[0]{BTZ$_{\rm M}$}
\newcommand{\ads}[1]{{\rm AdS}_{#1}}
\newcommand{\ds}[1]{{\rm dS}_{#1}}
\newcommand{\dS}[1]{{\rm dS}_{#1}}
\newcommand{\eds}[1]{{\rm EdS}_{#1}}
\newcommand{\sph}[1]{{\rm S}^{#1}}
\newcommand{\gn}[0]{G_N}
\newcommand{\SL}[0]{{\rm SL}(2,R)}
\newcommand{\cosm}[0]{R}
\newcommand{\hdim}[0]{\bar{h}}
\newcommand{\bw}[0]{\bar{w}}
\newcommand{\bz}[0]{\bar{z}}
\newcommand{\be}{\begin{equation}}
\newcommand{\ee}{\end{equation}}
\newcommand{\bea}{\begin{eqnarray}}
\newcommand{\eea}{\end{eqnarray}}
\newcommand{\pat}{\partial}
\newcommand{\lp}{\lambda_+}
\newcommand{\bx}{ {\bf x}}
\newcommand{\bk}{{\bf k}}
\newcommand{\bb}{{\bf b}}
\newcommand{\BB}{{\bf B}}
\newcommand{\tp}{\tilde{\phi}}
\hyphenation{Min-kow-ski}

\newcommand{\pa}{\partial}
\newcommand{\eref}[1]{(\ref{#1})}

\def\apr{\alpha'}
\def\str{{str}}
\def\lstr{\ell_\str}
\def\gstr{g_\str}
\def\Mstr{M_\str}
\def\lpl{\ell_{pl}}
\def\Mpl{M_{pl}}
\def\varep{\varepsilon}
\def\del{\nabla}
\def\grad{\nabla}
\def\tr{\hbox{tr}}
\def\perp{\bot}
\def\half{\frac{1}{2}}
\def\p{\partial}
\def\perp{\bot}
\def\eps{\epsilon}

\newcommand{\BC}{\mathbb{C}}
\newcommand{\BR}{\mathbb{R}}
\newcommand{\BZ}{\mathbb{Z}}
\newcommand{\bra}[1]{\langle{#1}|}
\newcommand{\ket}[1]{|{#1}\rangle}
\newcommand{\vev}[1]{\langle{#1}\rangle}
\newcommand{\Real}{\mathfrak{Re}}
\newcommand{\Imag}{\mathfrak{Im}}
\newcommand{\talpha}{{\widetilde{\alpha}}}
\newcommand{\Ham}{{\widehat{H}}}
\newcommand{\al}{\alpha}
\newcommand\x{{\bf x}}
\newcommand\y{{\bf y}}

\def\NPB{{\it Nucl. Phys. }{\bf B}}
\def\PL{{\it Phys. Lett. }}
\def\PRL{{\it Phys. Rev. Lett. }}
\def\PRD{{\it Phys. Rev. }{\bf D}}
\def\CQG{{\it Class. Quantum Grav. }}
\def\JMP{{\it J. Math. Phys. }}
\def\SJNP{{\it Sov. J. Nucl. Phys. }}
\def\SPJ{{\it Sov. Phys. J. }}
\def\JETPL{{\it JETP Lett. }}
\def\TMP{{\it Theor. Math. Phys. }}
\def\IJMPA{{\it Int. J. Mod. Phys. }{\bf A}}
\def\MPL{{\it Mod. Phys. Lett. }}
\def\CMP{{\it Commun. Math. Phys. }}
\def\AP{{\it Ann. Phys. }}
\def\PR{{\it Phys. Rep. }}

\renewcommand{\thepage}{\arabic{page}}
\setcounter{page}{1}

\rightline{hep-th/0401028}
\rightline{VPI-IPPAP-04-01}

\vskip 0.75 cm
\renewcommand{\thefootnote}{\fnsymbol{footnote}}
\centerline{\Large \bf WHAT IS QUANTUM THEORY OF GRAVITY?}
\vskip 0.75 cm

\centerline{{\bf Djordje Minic and
Chia-Hsiung Tze
}}
\vskip .5cm
\centerline{\it Institute for Particle Physics and Astrophysics,}
\centerline{\it Department of Physics, Virginia Tech}
\centerline{\it Blacksburg, VA 24061, U.S.A.}
\centerline{\it E-mail: dminic@vt.edu, kahong@vt.edu}
\vskip .5cm

\setcounter{footnote}{0}
\renewcommand{\thefootnote}{\arabic{footnote}}

\begin{abstract}
We present a line by line derivation of canonical quantum mechanics
stemming
from the compatibility of the statistical geometry of
distinguishable observations with the canonical
Poisson structure of Hamiltonian dynamics.
This viewpoint can be naturally extended to provide
a conceptually novel, non-perturbative formulation of quantum gravity. Possible observational implications of this new
approach are briefly mentioned.
\end{abstract}

\section{What is Quantum Mechanics?}
In this talk we would like to 1) give a line by line derivation of canonical quantum mechanics (QM) founded
on the compatibility of the statistical geometry of
distinguishable observations with the canonical
Poisson structure of Hamiltonian dynamics and 2) describe a
natural extension of this viewpoint so as to provide
a conceptually novel approach to the problem
of a non-perturbative formulation of quantum gravity, one which should reduce in the correspondence limit to general relativity (GR)
coupled to matter degrees of freedom. The presentation is based on our recent work \cite{tzeminic}.

To understand the fundamental structure of QM we reason
as follows: Assume that individual quantum events are 
statistical and statistically distinguishable (Postulate I). (This premise is of course a huge
conceptual leap in comparison to classical physics, but it is absolutely
crucial for the structure of QM.)
On the space of probability distributions there is a
natural metric, called Fisher metric, which provides a geometric measure of statistical 
distinguishability \cite{wootters}
\begin{equation}
ds^2=\sum_i \frac{dp_i^2}{p_i}, \quad \sum_i p_i =1, \quad p_i \geq 0.
\end{equation}
(This distance naturally arises as follows: to estimate probabilities
$p_i$ from frequencies $f_i$, given $N$ samples, when $N$ is large, the central limit theorem states that the probability for the
frequencies is given by the
Gaussian distribution
$\exp(-\frac{N}{2}\frac{(p_i-f_i)^2}{p_i})$. Thus a probability 
distribution $p^{(1)}_i$ can be distinguished from a given
distribution $p_i$ provided the
Gaussian $\exp(-\frac{N}{2}\frac{(p^{(1)}_i-p_i)^2}{p_i})$ is small. Hence
the quadratic form $\frac{(p^{(1)}_i-p_i)^2}{p_i}$, or its infinitesimal
form $\sum_i \frac{dp_i^2}{p_i}$, the Fisher distance, is the natural measure of
distinguishability.)
Now, upon setting $p_i = x_i^2$, making $p_i$ manifestly non-negative,
the Fisher distance reads $ds^2= \sum_i dx_i^2$ with $\sum x_i^2 =1$ i.e. the Euclidean metric on a sphere. Therefore 
the latter distance is nothing but the shortest distance along this unit sphere \cite{wootters}
\begin{equation}
\label{fisher}
ds_{12} = cos^{-1}(\sum_i \sqrt{p_{1i}} \sqrt{p_{2i}}).
\end{equation}

Next, we demand that on this metric space of probabilities one can define a canonical
Hamiltonian flow (Postulate II). So the dimensionality of such a symplectic space of $x_i$ must
be $even$ (hence, $\sum x_i^2 =1$ defines an odd-dimensional sphere). Then the Hamiltonian flow is given (locally) as
\begin{equation}
\frac{df(x_i)}{dt} = \omega_{ij}\frac{\partial h(x_i)}{\partial x_i}\frac{\partial f(x_i)}{\partial x_j} \equiv \{h,f\},
\end{equation}
where $\omega$ is a closed non-degenerate 2-form.
The compatibility of the symplectic form $\omega$ and the metric $g$
allows for the introduction of an almost complex structure $J \equiv \omega g^{-1}$
(in the matrix notation), $J^2 =-1$ (since the compatibility demands $\omega_{ij} g^{jk} \omega_{kl} = g_{il}$). Given this constant complex structure we may introduce complex coordinates on this space $\psi_a$ (and their conjugates $\psi_a^{*}$), so that
$\sum_i x_i^2 \equiv \sum_a \psi_a^{*} \psi_a =1$,
and thus $p_a = \psi_a^{*} \psi_a$.
This statistical distance is invariant under $\psi \to e^{J \alpha} \psi$, $J$ being the above integrable almost
complex structure.
Thus $\psi$ can be identified with $e^{J \alpha} \psi$.
Indeed an odd dimensional sphere can be viewed as a $U(1)$ Hopf -fibration of a complex
projective space $CP(n)$, a coset space
$\frac{U(n+1)}{U(n) \times U(1)}$.  $CP(n)$ is a homogeneous, isotropic and
simply connected Kahler manifold with a constant holomorphic sectional
curvature. 
The unique metric on $CP(n)$ is the Fubini-Study (FS) metric (which is but the above
statistical Fisher metric up to a multiplicative constant,
the Planck constant $\hbar$).  In the Dirac notation (using (\ref{fisher})
and the derived Born rule, $p_a = \psi_a^{*} \psi_a$), this FS metric reads :
\begin{equation}
\label{fs}
ds_{12}^2=4(cos^{-1}|\langle \psi_1|\psi_2\rangle|)^2 = 4(1 -
|\langle \psi_1|\psi_2 \rangle|^2)\equiv 4(\langle d\psi|d\psi\rangle
- \langle d\psi|\psi\rangle\langle \psi|d\psi\rangle),
\end{equation}
Thus $CP(n)$ is the underlying manifold of statistical events with a well defined Hamiltonian flow and as such provides a kinematical background on which a Hamiltonian
dynamics is defined.
The only Hamiltonian flow compatible with the isometries of $CP(n)$
(which are the unitaries $U(n+1)$) is given by a quadratic function of
$x_i$ or, alternatively, a quadratic
form in the pair $q_a \equiv Re(\psi_a)$ and $p_a \equiv Im(\psi_a)$), $h = {1 \over 2} \sum_a [ (p^a)^2 + (q_a)^2 ] \omega_a$, or in the usual notation, ${h = \langle \hat{H}\rangle}$, $\omega_a$ being the eigenvalues
of $\hat{H}$.
The Hamiltonian equations for the $\psi_a$ and its conjugate then give
the linear evolution equation (Schr\"{o}dinger equation), 
$
{d p_a \over dt} = \{h, p_a \}, \quad {d q^a \over dt} = \{h, q^a\}
$, that is $J \frac{ d|\psi\rangle}{dt} = H |\psi\rangle$.
Any observable, consistent with the isometries of the underlying space of events, is given as a quadratic function in the $q_a, p_a$. These are just the
usual expectation values of linear operators.

Everything we know about quantum mechanics is in fact contained in the 
geometry of $CP(n)$ \cite{landsman}, \cite{geomqm}:
entanglements come from the embeddings of the products of two complex projective
spaces in a higher dimensional one; 
geometric phases stem from the symplectic structure of $CP(n)$,
quantum logic, algebraic approaches to QM etc, are all contained in the geometric
and symplectic structure of complex projective spaces \cite{landsman}, \cite{geomqm}.
(While we only consider here the finite dimensional
case, the same geometric approach is extendable to generic infinite dimensional quantum mechanical
systems, including field theory.)
Finally, 
the following three lemmas are important for the material of the next section:
(A) The Fisher-Fubini-Study quantum metric (\ref{fs}) in the $\hbar \to 0$ limit becomes a spatial metric provided the configuration space for the quantum system under
consideration {\it is} space. For example, consider a particle moving in 3-dimensional Euclidean space. Then the
quantum metric for the Gaussian coherent state
$
\psi_l(x) \sim \exp(- \frac{{({\vec{x}}-{\vec{l}})}^2}{\delta l^2})
$
yields the natural metric in the configuration space,
in the $\hbar \to 0$ limit,
$
ds^2 = \frac{d {\vec{l}}^2}{\delta l^2}.
$
(B) Similarly, the time parameter of the evolution equation can be related to the quantum metric via 
\begin{equation}
\hbar ds = \Delta E dt, \quad \Delta E ^2 \equiv  \langle \psi H^2 \psi \rangle  - 
(\langle \psi H \psi \rangle)^2
\end{equation}
(C) Finally, the Schrodinger equation can be viewed as a geodesic equation on a $CP(n)= \frac{U(n+1)}{U(n) \times U(1)}$
\begin{equation}
{d u^a \over ds} + \Gamma^{a}_{bc} u^b u^c =
\frac{1}{2 \Delta E}Tr(H F^a_b) u^b.
\end{equation}
Here $u^a = \frac{d z^a}{ds}$ where $z^a$ denote the complex
coordinates on $CP(n)$, 
$\Gamma^{a}_{bc}$ is the connection obtained from the FS metric, and $F_{ab}$ is the canonical curvature 2-form valued in the holonomy gauge group $U(n)\times U(1)$.

The above geometric structure describing canonical QM, beautifully tested in numerous
experiments, is also very robust from the purely geometric point of view \cite{tzeminic}.
A consistent generalization of QM would doubtlessly be interesting from both
the experimental and theoretical viewpoints. 
Unlike various generalizations proposed in the past (which in many instances have lead to difficult conceptual problems) the one put forward in \cite{tzeminic}
extends the kinematical structure so that it is compatible with the
generalized dynamical structure! {\it The quantum symplectic and metric structure,
and therefore the almost complex structure become fully dynamical.}
The underlying physical reason for such a more general dynamical framework of the
above geometric formalism is found in the need for a quantum version
of the equivalence principle, a fundamental physical underpinning 
of a non-perturbative formulation
of quantum theory of gravity.

\section{And What is Quantum Theory of Gravity?}

The main intuition behind a quantum version of the 
equivalence principle is to demand the validity of the canonical QM, as laid out in the previous section, in every local neighborhood, {\it in the space of quantum events}. Here we envision a larger geometric structure whose tangent
spaces, viewed as vector spaces by definition, are just canonical Hilbert spaces.
(In this picture \cite{tzeminic} the tangent {\it spatial} transverse metric emerges from 
the {\it quantum} metric,
as in the lemma (A), by assuming that the underlying configuration space is space
and time appears as a measure of the geodesic distance in this general
space of statistical events (the events do not have to be necessarily
distinguishable!), as in lemma (B). Finally, the longitudinal 
spatial coordinate corresponds to the dimensionality of
the tangent Hilbert spaces.) The crucial point is to allow for any metric and symplectic form in
the geodesic version of the evolution equation
(lemma (C)). These are in turn determined dynamically \cite{tzeminic}.

Our postulates (I) and (II) as stated above can indeed be naturally extended
by allowing both the metric and symplectic form
on the space of quantum events to be no longer rigid but fully
dynamical entities.
In this process, just as in the case of spacetime in GR,  the space of quantum events becomes dynamical
and only individual quantum events make sense
observationally. 
Specifically, we do so by relaxing the first postulate to allow for {\it any} statistical (information) metric while
insisting on the compatibility of this metric with the symplectic structure
underlying the second postulate. Physics is therefore required to be diffeomorphism
invariant in the sense of information geometry  such
that the information geometric and symplectic structures remain compatible, requiring only a $strictly$ (i.e non-integrable) almost complex structure $J$.
This extended framework readily implies that the wave functions labeling the event space, while still unobservable, are no longer relevant. They are in fact as meaningless as coordinates
in GR.  There are no longer issues related to reductions of wave packets and associated measurement
problems. At the basic level of our scheme, there are only dynamical correlations of quantum events. The observables are furnished by diffeomorphism invariant quantities
in the space of quantum events.

To find the kinematical arena for this generalized framework we seek
as coset of $Diff(C^{n+1})$ such that locally the latter looks like $CP(n)$ and allows
for a compatibility of its metric and symplectic structures, expressed in
the existence of a (generally non-integrable) almost complex structure.
The following nonlinear Grassmannian
\begin{equation}
Gr(C^{n+1}) = Diff(C^{n+1})/Diff(C^{n+1},C^n \times \{0\}) ,
\end{equation}
with $n = \infty$ fulfills these requirements \cite{vizman}.   
$Gr(C^{n+1})$ is a nonlinear analog of a complex Grassmannian since it is the space of (real) co-dimension 2 submanifolds, namely a hyperplane $C^n \times [0]$ passing through the origin in $C^{n+1}$.  Its holonomy group $Diff(C^{n+1} , C^n\times \{0\})$ is the group of diffeomorphisms preserving the hyperplane  $C^n \times \{0\}$ in $C^{n+1}$. Just as $CP(n)$ is a co-adjoint orbit of $U(n+1)$, $Gr(C^{n+1})$ is a coadjoint orbit of the group of volume preserving diffeomorphisms of $C^{n+1}$. As such it is a symplectic manifold with a canonical Kirillov-Kostant-Souriau symplectic two-form $\Omega$ which is closed ($d\Omega=0$) but not exact. Indeed the latter 2-form integrated over the submanifold is nonzero; its de Rham cohomology class is integral.  This means that there is a principal 1-sphere, a $U(1)$ or line bundle over $Gr(C^{n+1})$ with curvature $\Omega$. This is the counterpart of the 
$U(1)$-bundle of $S^{2n+1}$ over $CP(n)$ of quantum mechanics. It is also known that there is an almost complex structure given by a 90 degree rotation in the two dimensional normal bundle to the submanifold.  While $CP(n)$ has an integrable almost complex structure and is therefore a complex manifold, in fact  a Kahler manifold, this is $not$ the case with $Gr(C^{n+1})$. Its almost complex structure $J$ is strictly $not$ integrable in spite of its formally vanishing Nijenhius tensor. While the vanishing of the latter implies integrability in the finite dimensional case, such a conclusion no longer holds in the infinite dimensional setting. However what we do have in $Gr(C^{n+1})$ is a strictly (i.e. non-Kahler) almost Kahler  manifold since there is by way of the almost structure $J$ a compatibility between the closed symplectic 2-form $\Omega$ and the Riemannian metric $g$ which $locally$ is given by $g^{-1}\Omega = J$. Clearly, it would be very interesting to understand how unique is the structure
of $Gr(C^{n+1})$.

Just as in standard geometric QM, the probabilistic interpretation should be found in the
definition of geodesic length on the new space of quantum states/events.
Notably since $Gr( C^{n+1})$ is only a strictly almost complex, its $J$ is only locally complex. This fact translates into the existence of only local time and local metric on the
space of quantum events. 
The local temporal
evolution equation is a geodesic equation on the space of quantum events
$
{d u^a \over d\tau} + \Gamma^{a}_{bc} u^b u^c =
\frac{1}{2 E_p}Tr(H F^a_b) u^b
$
where now $\tau$ is given through the metric
$\hbar d\tau = 2 E_p dt$, where $E_p$ is the Planck energy.
$\Gamma^{a}_{bc}$ is the affine connection associated with this general metric 
$g_{ab}$ and $F_{ab}$ is a general curvature 2-form in the holonomy gauge group $Diff(C^{n+1},C^n \times \{0\})$.
This geodesic equation follows from the conservation of the
energy-momentum tensor 
$
\nabla_a T^{ab} =0
$
with 
$
T_{ab} = Tr(F^{ac}g_{cd}F^{cb} -\frac{1}{4} g_{ab} F_{cd}F^{cd}
+ \frac{1}{2E_p}H u_a u_b).
$
Since both the metrical and symplectic data are also contained in
$H$ and are $\hbar \to 0$ limits of their quantum counterparts \cite{geomqm}, \cite{tzeminic}, we have here a non-linear ``bootstrap'' between the space of
quantum events and the generator of dynamics.
The diffeomorphism invariance of the new quantum phase space
is explicitly taken into account in the following dynamical scheme 
\cite{tzeminic}:
\begin{equation}
\label{BIQM1}
R_{ab} - \frac{1}{2} g_{ab} R  - \lambda g_{ab}= T_{ab}
\end{equation}
($\lambda = \frac{n+1}{\hbar}$ for $CP(n)$; in that case $E_p \to \infty$).
Moreover we demand for compatibility
\begin{equation}
\label{BIQM2}
\nabla_a F^{ab} = \frac{1}{E_p} H u^b.
\end{equation}
The last two equations imply via the Bianchi identity a conserved
energy-momentum tensor, $\nabla_a T^{ab} =0$ .  The latter, taken
together with the conserved ``current'' $j^b \equiv \frac{1}{2E_p} H u^b$,
i.e. $\nabla_a j^a =0$,
results in the generalized geodesic Schr\"{o}dinger equation. As in GR it will be crucial to understand both the local and global features of various
solutions to the above dynamical equations.
The kinematical structure of ordinary QM is compatible with
our general dynamical formulation and thus we naturally expect that our general formalism is
compatible with all known cases in which quantum theories of gravity have
been non-perturbatively defined, albeit in {\it fixed} asymptotic backgrounds
(such as string theory in asymptotically $AdS$ spaces).

What determines the form of the Hamiltonian $H$?
The only requirement is that $H$ should define a canonical quantum mechanical
system whose configuration space {\it is} space and whose dynamics defines
a consistent quantum gravity in an asymptotically flat background.
We are aware of only one example satisfying this criterion: 
Matrix theory \cite{matrix}!
{\it Thus our proposal defines a background independent, non-perturbative,
holographic formulation of Matrix theory} \cite{tzeminic}. 
This choice of the Hamiltonian might at first appear artificial given the generality of
our proposal.  Yet we
note that, from the suggestive geodesic form of the Schr\"{o}dinger equation, $H$ can be viewed as a ``charge'', and thus may be determined in a quantum theory of
gravity by being encoded in the non-trivial symplectic topology of the space of quantum events. 
Such a realization may well be possible here since our non linear Grassmannian is non-simply connected \cite{vizman}.

Finally, among the possible observational implications
of our proposal, those we are currently attempting to understand are:
1) possible deviations from the Planck law
involving primordial gravitational waves, or Hawking radiation;
2) deviations from the classic QM formula for the
vacuum energy (which underlies the cosmological constant problem);
3) relativity of the double slit experiment:
once we relax postulate I, so that any information metric is allowed,
the relativity (observer dependence) of canonical QM experiments (such as the double-slit
experiment) becomes possible;
4) highly constrained deviations from linearity (superposition principle), and canonical QM entanglement;
5) the fact that in our proposal the generalized geometric phase
is in $Diff(C^{n+1},C^n \times \{0\})$, 
is also amenable to experimental tests.

{\bf Acknowledgments:}
We thank Tom Banks, Raphael Bousso, Ram Brustein, Lay Nam Chang,
Marty Einhorn, Willy Fischler, Jim Hartle,
Bob McNees, Joan Simon and Tatsu Takeuchi for recent discussions
of our work.
{\small D.M.} would like to
thank the organizers of the QTS3 conference for
providing him with an opportunity to present this work. 
{\small DM} was supported in part by the U.S. Department of Energy under contract DE-FG05-92ER40677 and by NSF grant PHY-9907949 at KITP, Santa
Barbara, the stimulating atmosphere of which is gratefully acknowledged.

\end{document}